\documentclass[runningheads]{llncs}
\usepackage[T1]{fontenc}
\usepackage{dingbat}
\usepackage{graphicx}
\usepackage{url}
\usepackage{amsmath}
\usepackage{booktabs}
\usepackage{algpseudocode}
\usepackage{algorithm}
\usepackage{multirow}
\usepackage{subcaption} 
\begin{document}
\title{CODE ACROSTIC: Robust Watermarking for Code Generation}
%
%
\author{Li Lin\inst{1}  \and Siyuan Xin\inst{2}
Yang Cao\inst{3}\orcidID{0000-0002-6424-8633} \and
Xiaochun Cao\inst{4}\orcidID{0000-0001-7141-708X}}
\authorrunning{L. Author et al.}
%
\institute{Institude of Science Tokyo \\ \email{lin.l.ac@m.titech.ac.jp} \and
Shanghai University \\ \email{zeshixing@shu.edu.cn}
\and
Institude of Science Tokyo
\\ \email{cao@c.titech.ac.jp}\\
\and
Sun Yat-sen University,\\
\email{caoxiaochun@mail.sysu.edu.cn}}

\maketitle              
\begin{abstract}
Watermarking large language models (LLMs) is vital for preventing their misuse, including the fabrication of fake news, plagiarism, and spam. 
It is especially important to watermark LLM-generated code, as it often contains intellectual property.
However, we found that existing methods for watermarking LLM-generated code fail to address \textit{comment removal attack}.
In such cases, an attacker can simply remove the comments from the generated code without affecting its functionality, significantly reducing the effectiveness of current code-watermarking techniques.
On the other hand, injecting a watermark into code is challenging because, as previous works have noted, most code represents a low-entropy scenario compared to natural language. 
Our approach to addressing this issue involves leveraging prior knowledge to distinguish between low-entropy and high-entropy parts of the code, as indicated by a \textit{Cue List} of words.
We then inject the watermark guided by this Cue List, achieving higher detectability and usability than existing methods.
We evaluated our proposed method on
\textit{HumanEval}
and compared our method with three state-of-the-art code watermarking techniques. 
The results demonstrate the effectiveness of our approach.

\end{abstract}

\section{Introduction}
\label{sec:intro}
The rapid proliferation of Large Language Models (LLMs) has revolutionized various fields, showcasing their potential in automating tasks \cite{boiko2023autonomous}, generating creative content \cite{zhao2023survey}, and solving complex problems \cite{li2023starcoder}. However, the growing reliance on LLMs has also raised significant concerns, particularly regarding the potential for plagiarism and dishonesty \cite{dishonesty}, especially in contexts where the originality of content is crucial, such as homework programming \cite{ghimire2024coding} or open-source activities. 
Furthermore, AI hallucination would lead to misinformation and degradation of future language models \cite{hull}.
This issue of ownership and accountability remains an open problem. 
Tools are formulated to counteract the plagiarism associated with LLMs, part of them are the Post-hoc detection approaches \cite{zhong_neural_2020,koike_outfox_2024} (i.e. without changing the text generation process), which analyze the features of generated context and judge the genuineness across different samples. 
However, after the machine-originated text turned out to be flawlessly close to the human-written one, as \textit{the first sentence is generated by one of the LLMs}, the Post-hoc fashion approaches are expected to struggle. 
Watermark-based methods have been developed as a potential solution to address challenges in tracking and verifying content, especially in programming languages.

\begin{figure}[t]
    \centering
    \includegraphics[width=.8\linewidth]{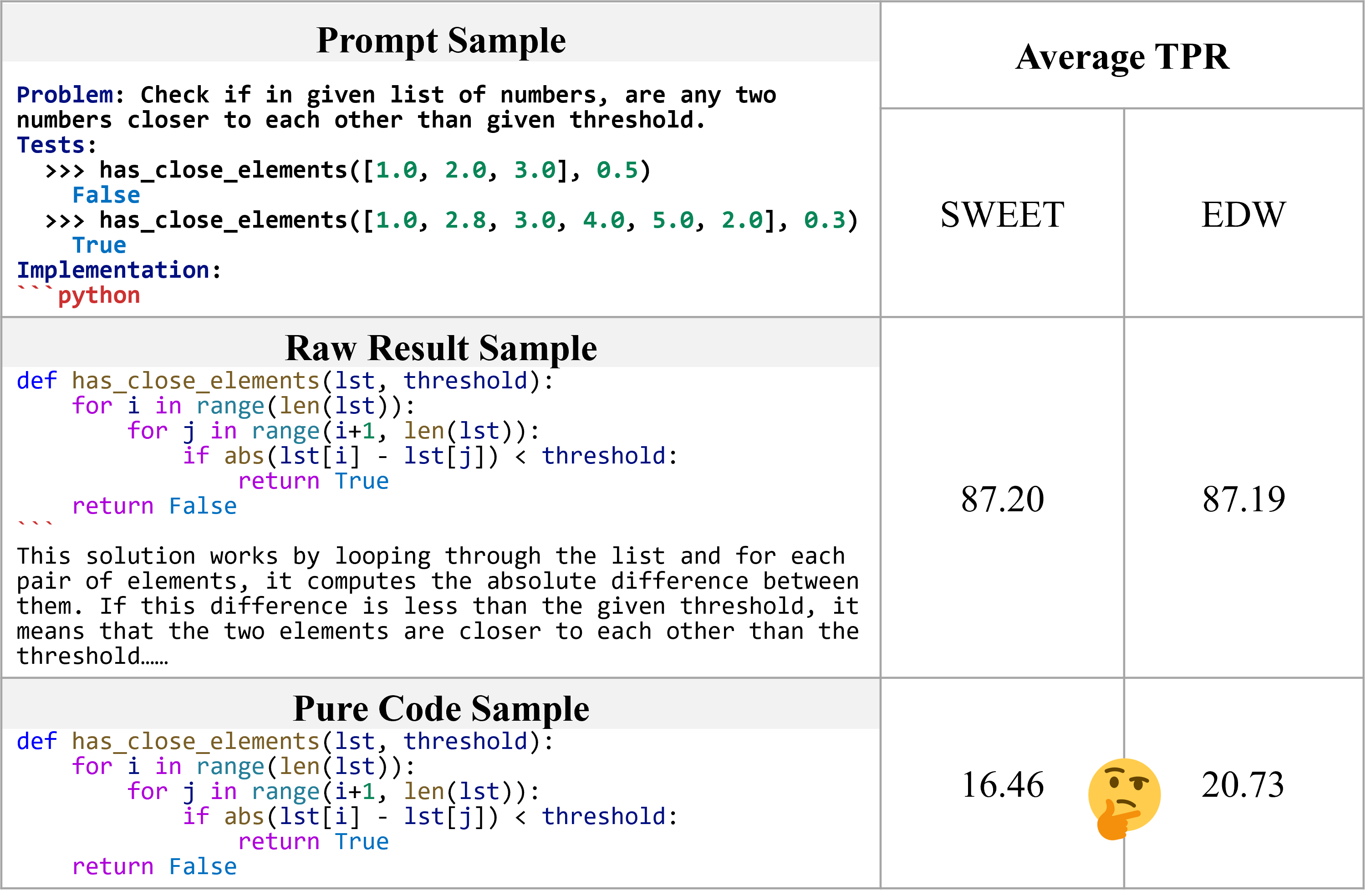}
    \caption{\textit{Comment Removal Attack} that simply removes the comments from the generated code without affecting its functionality(True Positive Rate on HumanEval Dataset here), can significantly reduce the effectiveness of the state-of-the-art code-watermarking techniques \textbf{EWD} and \textbf{SWEET}.}
    \label{fig:enter-label}
\end{figure}

Watermarking techniques have evolved significantly in recent years, beginning with the approach introduced by Kirchenbauer(\textbf{KGW}) \cite{kirchenbauer_watermark_2023}, where service providers embed identifiable marks or patterns into content generated by LLMs for future tracking and origin verification. 
Building on this, \textbf{SWEET} \cite{lee_who_2024} enhances watermarking by applying an entropy threshold, selectively injecting watermarks into high-entropy regions while clamping low-entropy tokens during the detection phase. 
This architecture results in a more robust watermarking scheme. 
Later, \textbf{EWD} \cite{lu2024entropy} further refines this by incorporating entropy into the weighting of the detection process, producing a "soft" detector. 



However, existing watermarking methods \cite{kirchenbauer_watermark_2023,lee_who_2024,lu2024entropy} for code generation neglect \textit{robustness} the watermark against attacks.
As illustrated in Figure \ref{fig:enter-label}, a simple but effective comment removal attack can degrade the detectability of the watermark to a level that is unacceptable.  

To tackle the challenges of watermarking in code generation, we introduce a novel approach called \textbf{Code Acrostic}, to improve robustness and detectability. 
The key insight behind Code Acrostic is that some segments of the code can tolerate modifications without affecting functionality. 
Leveraging this, Code Acrostic selectively injects watermarks into areas resilient to changes and minimally disruptive to the code’s correctness. 
The core of this method is the use of high-entropy tokens, which are less predictable and more adaptable to modifications, making them ideal for watermarking. 
By pre-calculating a \textbf{Cue List}, which identifies the tokens immediately \textbf{preceding} these high-entropy tokens, Code Acrostic inserts watermark tokens only \textbf{after} the cue tokens. 
This creates a sparse watermark that enhances detectability and robustness while minimizing the impact on the usability of the code, ensuring it remains fully functional.

\textbf{Summary of our technical contributions}. 
For the first time, we identify that comment removal attacks present an overlooked challenge for existing code watermarking methods. 
Second, we introduce Code Acrostics, a novel watermarking technique for LLM-generated code that maintains high detectability, even against comment removal attacks. 
Finally, we experimentally evaluated the robustness, detectability, and usability of code watermarking methods, confirming the effectiveness and superiority of our proposed approach.

\section{Related Work}

\label{sec:related_works}




\subsection{Watermark-Based Methods}
Watermark techniques are considered to be important to counter the abuse of LLM \cite{liu2024survey}, and it is also a tool with abundant research history \cite{atallah_natural_2001}, focused on embedding identifiers in natural language for tracking and attribution purposes. 

Kirchenbauer et al. \cite{kirchenbauer_watermark_2023} formulate the framework for watermarking in LLM-generated text, introducing a groundwork that integrates watermarking directly into the generation process. Christ et al. \cite{christ_undetectable_2023} introduced advanced cryptographic techniques to enhance watermark security, making detection more robust under attack or modification. Fairoze et al. \cite{fairoze2023publicly} introduce the important idea of public-key cryptography to the scheme of watermark, which is able to be used at the proof of the possession of a LLM or detection when it is required. Furthermore, Fu et al. \cite{fu_gumbelsoft_2024} refined the watermark injection process using the Gumbel-max trick, improving both the imperceptibility and detectability of watermarks. Furthermore, SynthID\cite{dathathri2024scalable} presents a random mechanism called tournament sampling for the watermark injection, which alleviates the destruction to the generation quality. For the situation that the model cannot be modified, Yang et al. \cite{yang2023watermarking} explored injecting watermarks via bi-directional models in a black-box setting, addressing scenarios where the generation pipeline is inaccessible.

\subsection{Watermarking for Code Generation}

The application of watermarking in code generation presents unique challenges, as the structure and functionality of code must remain untouched. Li et al. \cite{li2024resilient} introduced ACW, which uses a set of predefined transformations to inject watermarks without compromising code correctness.
By modifying the training dataset, Sun et al. \cite{sun2023codemark} proposed injecting watermarks effectively, biasing the model to generate watermarked code.
Guan et al. \cite{guan2024codeip} leveraged code syntax engines and type predictors to embed watermarks in a way that aligns with the syntactical and semantic properties of the code. Likely, Kim et al. \cite{kim2025marking} present a framework called STONE, which selectively inject watermark to non-syntax tokens.
Lee et al. \cite{lee_who_2024} utilized selective entropy thresholds to embed watermarks in high-entropy regions, while Lu et al. \cite{lu2024entropy} incorporated entropy as part of the detection weight. Furthermore, Suresh et al. \cite{suresh2024is} evaluate the robustness of existing methods on watermarking code and point out that vulnerability to attack is a common issue for the current methods.

Despite these advancements, a critical issue remains unsolved: the entropy of LLM-generated code often differs significantly from that of real-world code. This mismatch renders many existing methods vulnerable to attacks, such as comment removal attack, where low-entropy regions like comments are stripped away without affecting functionality. Addressing this vulnerability requires techniques that adaptively identify and watermark high-entropy regions, ensuring robustness even under adversarial conditions.

\section{Preliminaries}
\label{sec:preliminaries}




\subsection{Watermark for LLMs}
The watermark technique for LLM can be viewed as two split processes, watermark injection and watermark detection. A typical scenario is that the LLM provider injects the watermark into the text which they provide to the customer, and provides the detection toolkit for the customer if needed.



\subsubsection{Watermark Injection}
Watermarking techniques require a fundamental hash function that splits the token space into a green list and a red list \cite{kirchenbauer_watermark_2023}. 
In natural language, the green-to-red list ratio is expected to be balanced as long as the hash is random. The goal of watermarking methods is to alter this balance by introducing a bias to the green list. To enhance a specific green list token's likelihood by a magnitude \( \delta \), 
\begin{equation}
    l_{wt}' =l_{wt} + \delta F_{hash}(w_t, w_p,\gamma,\theta_{key}).
\end{equation}
\subsubsection{Watermark Detection}
For detection, the text is first tokenized using the same tokenizer, and the length of the sequence is denoted as $T$, while the hit rate of the green list is calculated as $|s|_G$, as
\begin{equation}
    |s|_G = \sum_{i=0}^{t}w_i \in V_G(w_p,\gamma,\theta) .
\end{equation}
Then, a z-statistic under null-hypothesis is applied to evaluate how far this ratio deviates from the expected natural ratio: 
\begin{equation}
\label{z_score}
    z=\frac{|S|_G-\gamma T}{\sqrt{T\gamma(1-\gamma)}}.
\end{equation}
If the z-score exceeds a predetermined threshold, known as the \( z\text{-threshold} \), it indicates with high confidence that the text was generated by an LLM.

\section{Our Method}
\label{sec:Methodologies}
The proposed method, \textbf{Code Acrostic}, employs a three-step approach: the construction of a \textbf{Cue List}, the selective watermark injection through \textbf{Cue Sampling}, and the final identification using \textbf{Cue Detection}. 

\subsection{Cue List}

The \textbf{Cue List} is a central element of Code Acrostic, serving as a filter to determine high-entropy regions suitable for watermark injection. By focusing on tokens with high entropy, this method avoids disrupting low-entropy sections, such as comments or predictable code patterns, which are more vulnerable to removal or modification. Importantly, the Cue List must remain confidential to ensure robustness against adversarial attacks.

To compute the Cue List, Code Acrostic uses a 1-gram model, which calculates the conditional probability of the next token based on a single preceding token. By introducing a third-party corpus, such as \cite{partialPython}, denoted as \( C = (t_1, t_2, \dots, t_N) \) with size \( N \), we calculate the 1-gram probabilities. The corpus \( C \) is arbitrarily chosen, and any alternative can be used without restriction. The function \( k_{t_a}(t_b) \) represents the number of times the pair \((t_a, t_b)\) appears consecutively in a sequence of pairs \((t_i, t_{i+1})\), where \(i\) ranges from 0 to \(N-1\). For any token pair \((t_a, t_b)\), the co-occurrence frequency \( k_{t_a}(t_b) \) is defined as:

\begin{equation}
    k_{t_a}(t_b)=\sum_{i=0}^{N-1}(\{t_a,t_b\} = \{t_i,t_{i+1}\}) .
\end{equation}

Using these frequencies, a list of probabilities for token pairs is
\begin{equation}
    K= \{k_{t_0}(T),k_{t_1}(T),\cdots,k_{t_N}(T)\}.
\end{equation}

It gets the Cue List $\hat{K}$, by:
\begin{equation}
    \forall k \in \hat{K} ,k \in K \And k'=\frac{k}{\sum k}
\end{equation}
And formulate the cue poll with a threshold $\beta$, as :

\begin{equation}
F_{\text{cue}}(k) = \left\{ \begin{array}{cl}
1 & : -\sum k'*\log k' < \beta \\
0 & : \text{else}
\end{array} \right.
\end{equation}

This thresholding ensures that only high-entropy tokens, which are tokens with greater unpredictability, are included in the Cue List. Such tokens are more robust against modifications and less likely to affect code correctness.

\subsection{Cue Sample}
 \begin{algorithm}[H]
\caption{Code Acrostics watermarks generation}
\label{algo:injection}
\begin{algorithmic}[1]
\Function {Generate\_watermark}{$w_{prompt}$}
    \State $t_{\text{previous}} \gets w_{prompt}[-1]$
    \While{ $ t_{\text{previous}} \neq t_{EOS}$ }
    \State  $l \gets F_{llm}(w_{prompt})$
    \State  $hash \gets \delta * H(\gamma, w_{prompt}[-N:-1])$
    \State $Cue \gets F_{\text{cue}}(t_{\text{previous}})$
    \State $w_{prompt} \gets w_{prompt} + Sampler(Cue*hash)$

    \EndWhile
    \State \Return $w_{prompt}$
\EndFunction
\end{algorithmic}
\end{algorithm}

Cue Sampling leverages the Cue List to selectively inject watermarks into sequences during code generation. By focusing on regions identified as high-entropy, the watermarking process minimizes its impact on usability while maximizing robustness and detectability. With the Cue List, we can sample the original watermark function, as
\begin{equation}
    w_t' = F_{\text{cue}} \circ F_{\text{sample}} \circ F_{llm}(w_{1:t-1}).
\end{equation}
This cue function is a component of the watermark function, and can be viewed as a high-order sampler over the $F_{\text{sample}}$. The algorithm \ref{algo:injection} showcases the pseudo-code of watermark injection. 

\subsection{Cue Detection}

Code Acrostic exclusively detect tokens next to the token from Cue List by adopting the Cue List as a token filter. Given a sequence to be detect denoted as $W' = (w'_0,w'_1,\cdots,w'_t)$, the counted token number $T$ and green list token $|S|_G$ should be calculated as:
\begin{equation}
    \begin{split}
|S|_G=\sum_{i = 1}^{t}F_{cue}(w'_{i-1})F_{hash}(w'_i,w'_{i-N:i-1},\gamma,\theta_{key})
    \end{split}
\end{equation}

As the process described with equation \ref{z_score}, final z-score can be calculated and apply to a comparison with the z-threshold. The algorithm \ref{algo:detection} showcases this process. The detection phase involves verifying the presence of watermarks by analyzing tokens adjacent to those in the Cue List. By using the same threshold and hash function as in the sampling phase, the detection process ensures consistency and accuracy.

\begin{algorithm}[t]
\caption{Code Acrostic Watermarks Detection}
\label{algo:detection}
\begin{algorithmic}[1]
    \Function{Detect\_Watermarks}{$w_{test}$}
    \State $S \gets 0 \And T \gets 0$
    \ForAll {$i=2,3,\cdots,|w_{test}|$}
        \State $t_{\text{previous}} \gets w_{test}[i-1]$
        \If {$F_{\text{cue}}(t_{\text{previous}})$}
            \State $T \gets T + 1$
            \State $G \gets H(\gamma, w_{test}[i-N:i-1])$
            \If{$G[W_{test}[i]]$} \Comment{if the token is green}
                \State $S \gets S + 1$
            \EndIf
        \EndIf
    \EndFor
    \State $z_{score} = \frac{S -\gamma T}{\sqrt{T\gamma(1-\gamma)}}$
    \State \Return $z_{score} > z_{\text{threshold}}$
    \EndFunction
\end{algorithmic}
\end{algorithm}


\section{Experiments and Benchmarks}

\begin{table*}[t]
\centering
\begin{tabular}{lccc}
\toprule
\multirow{2}{*}{Methods}               & \multicolumn{3}{c}{Human Eval}         \\ \cmidrule{2-4} 
                                       & Pass@1 $\uparrow$ & AUROC $\uparrow$ &   \(\Delta\) AUROC (\%) over KGW                                   \\ \midrule
Raw                                    & 53.05             & -                & -                                    \\
Code Acrostic (Ours)                   & 51.20             & \textbf{0.772}   & \textbf{+12.70\%}       \\ 
KGW \cite{kirchenbauer_watermark_2023} & 54.27             & 0.685            & -                                    \\
SWEET \cite{lee_who_2024}              & \textbf{54.88}    & 0.692            & +1.02\%          \\ 
EWD \cite{lu2024entropy}               & \textbf{54.88}    & 0.750            & +9.49\%           \\ \bottomrule
\end{tabular}
\caption{\textbf{Detectability and usability} on Human Eval Dataset comparing to other major approaches. "Raw" represents the original output of LLM without any watermark technique. Improvements in AUROC are calculated relative to KGW. Green indicates positive improvement.} 
\label{Main_result}
\end{table*}

\label{sec:experiment}
\subsection{Setups}
We evaluates the the \textbf{robustness},\textbf{ detectability and usability} of the proposed method. 

\textbf{Implementaion.} For a fair comparison, the experiments leverage the \textit{MarkLLM} framework \cite{pan2024markllm}, implemented with the DeepSeek-Coder-1.3b model \cite{guo2024deepseek}. Code Acrostic and other competing approaches were integrated into this infrastructure. The source code for this study is publicly available at \url{https://github.com/xhaughearl/code_acrostic}, ensuring reproducibility and transparency.  

\textbf{Dataset.} The \textit{HumanEval} benchmark \cite{chen2021codex}, designed to evaluate code generation models, was used for assessing the performance of watermarking techniques. This benchmark consists of 164 coding problems, each accompanied by test cases to evaluate code correctness.  

\textbf{Metrics.} To evaluate the proposed method, we assess performance using \textbf{Pass@1}, which measures whether the generated code passes all test cases (total 164 problems for $HumanEval$\cite{chen2021codex}) on the first attempt. 
For watermarking, we evaluate the \textbf{True Positive Rate (TPR)}, which reflects the accuracy of detecting watermarked code,While the False Positive Rate(FPR) is similar metric on Non-marked code, as:

\begin{equation}
 \text{TPR} = \frac{\text{Marked Code Detected}}{\text{All Marked Code}}; 
\end{equation}
while False Positive Rate (FPR) evaluates the proportion of raw (non-marked) code that is mistakenly classified as watermarked as 
\begin{equation}
 \text{FPR} = \frac{\text{Raw Code Detected}}{\text{All Raw Code}}. 
\end{equation}

We also use the \textbf{Area Under the Receiver Operating Characteristic Curve (AUROC)} to gauge the detector's ability to distinguish between watermarked and non-watermarked code across varying thresholds $t$, as:
\begin{equation}
    \text{AUROC} = \int_{-\infty}^{\infty} \text{TPR}(t) , 
d(\text{FPR}(t)). 
\end{equation}
These metrics provide a holistic view of both the functionality of the watermarked code and the effectiveness of the watermark detection process.  

\subsection{Robustness}
\begin{table*}[t!]
\centering
\begin{subtable}[t]{0.48\linewidth}
\centering
\begin{tabular}{cc|cccc}
\toprule
Natural & Code & CA    & KGW   & SWEET& EWD \\
\midrule
\checkmark & \checkmark & 85.37 & 24.39 & \textbf{87.20} & 87.19 \\
 & \checkmark & \textbf{34.76} & 17.07 & 16.46 & 20.73 \\
\bottomrule
\end{tabular}
\caption{\textbf{Detectability}. TPR of each approach w/ or w/o natural language (comments) mixed, with a uniform z-threshold.}
\label{Purification}
\end{subtable}\hfill
\begin{subtable}[t]{0.48\linewidth}
\centering
\begin{tabular}{l|cccc}
\toprule
$\delta$ & 0 & 8 & 12 & 40 \\
\midrule
Pass@1 & \textbf{53.05} & 50.00 & 37.80 & 15.85 \\
TPR & & 37.20 & \textbf{59.15} & 20.73 \\
\bottomrule
\end{tabular}
\caption{Parameters Ablation. Different $\delta$ with Pass@1 and TPR metrics.}
\label{parameters}
\end{subtable}
\caption{Comparison of detectability and parameter ablation results.}
\label{tab:combined}
\end{table*}



It compares the results before and after filtering out the natural language portion from the detection scope. It shows that all methods exhibit higher detectability when natural language is included in the scope. However, once the natural language is filtered out, \textbf{the proposed method stands out with the highest detectability}. This suggests that the proposed method effectively embeds the watermark directly into the code itself, rather than relying on the presence of natural language. It is also notable that SWEET and EWD require longer runtimes compared to CA and KGW when detecting watermark with LLMs.

\subsection{Detectability and usability}
Detectability and usability are discussed in this section, and the proposed method achieves superior detectability with acceptable usability.

\textbf{usability.} Maintaining an acceptable level of code accuracy is essential for watermarked code. Table \ref{Main_result} shows how various watermarking methods impact accuracy. While our method experiences a slight degradation in Pass@1, others increase Pass@1 by about 1\%. This improvement is likely due to the watermark acting as a random sampler, introducing beneficial variability in the generated content. As noted by Kirchenbauer et al. \cite{kirchenbauer_watermark_2023}, mild watermarking schemes can enhance content quality. However, Code Acrostic does not exhibit this effect, likely due to its strict Cue List, which prioritizes robustness and detectability, limiting flexibility in watermark placement.

\textbf{Detectability. }
Table \ref{Main_result} showcases the results, highlighting that the proposed method achieves superior detectability on the HumanEval benchmark. The AUROC shows that Code Acrostic excels in detectability, thereby outperforming other methods. As for why the other approaches are significantly struggling, part of the reason is that the watermark is injected into the natural language part of the content.


\subsection{Ablation Experiments}


Different magnitudes of the watermark were tested, with the results for Pass@1 and TPR presented in Table \ref{parameters}. As expected, increasing the watermark's magnitude tends to trade off some accuracy, reflected in a decrease in Pass@1. As \cite{hu2024inevitable} pointed out, there is a trade-off between the detectability and the usability for the series of watermark methods.
Interestingly, we observed that the TPR initially increases as the watermark magnitude grows, but it starts to decline when the magnitude becomes too large. This pattern highlights the importance of carefully balancing watermark magnitude to optimize both accuracy and detectability.

\section{Conclusion}
\label{sec:page}
In this paper, we presented Code Acrostic, a novel approach to watermarking in programming languages. With the aid of a list of high-entropy token cues, Code Acrostic embedded watermarks in resilient, high-entropy code regions, and ensured minimal disruption to code \textbf{usability} while maintaining high \textbf{detectability}. Furthermore, we introduced a code-only measurement technique that eliminates the influence of natural language elements, thereby measuring the \textbf{robustness} of the watermark. The proposed method has demonstrated superior performance, and we hope this method offers new insights for further research in this field.
\vfill\pagebreak
\section{Acknowledgement}
This work is supported by JSPS KAKENHI JP23K24851, JST PRESTO JPMJPR23P5, JST CREST JPMJCR21M2, JSPS  JPJSBP120247401, JST NEXUS JPMJNX25C4. We also thank the anonymous reviewers and all who contributed to improving the quality of this paper.

\bibliographystyle{unsrt}
\bibliography{ijcai25}

@article{kim2025marking,
  title={Marking Code Without Breaking It: Code Watermarking for Detecting LLM-Generated Code},
  author={Kim, Jungin and Park, Shinwoo and Han, Yo-Sub},
  journal={arXiv preprint arXiv:2502.18851},
  year={2025}
}

@inproceedings{
suresh2024is,
title={Is Watermarking {LLM}-Generated Code Robust?},
author={Tarun Suresh and Shubham Ugare and Gagandeep Singh and Sasa Misailovic},
booktitle={The Second Tiny Papers Track at ICLR 2024},
year={2024},
url={https://openreview.net/forum?id=8PhI1PzSYY}
}

@article{dathathri2024scalable,
  title={Scalable watermarking for identifying large language model outputs},
  author={Dathathri, Sumanth and See, Abigail and Ghaisas, Sumedh and Huang, Po-Sen and McAdam, Rob and Welbl, Johannes and Bachani, Vandana and Kaskasoli, Alex and Stanforth, Robert and Matejovicova, Tatiana and others},
  journal={Nature},
  volume={634},
  number={8035},
  pages={818--823},
  year={2024},
  publisher={Nature Publishing Group UK London}
}

@article{liu2024survey,
  title={A survey of text watermarking in the era of large language models},
  author={Liu, Aiwei and Pan, Leyi and Lu, Yijian and Li, Jingjing and Hu, Xuming and Zhang, Xi and Wen, Lijie and King, Irwin and Xiong, Hui and Yu, Philip},
  journal={ACM Computing Surveys},
  volume={57},
  number={2},
  pages={1--36},
  year={2024},
  publisher={ACM New York, NY}
}

@article{zhao2023survey,
  title={A survey of large language models},
  author={Zhao, Wayne Xin and Zhou, Kun and Li, Junyi and Tang, Tianyi and Wang, Xiaolei and Hou, Yupeng and Min, Yingqian and Zhang, Beichen and Zhang, Junjie and Dong, Zican and others},
  journal={arXiv preprint arXiv:2303.18223},
  year={2023}
}

@article{li2024resilient,
  title={Resilient Watermarking for LLM-Generated Codes},
  author={Li, Boquan and Zhang, Mengdi and Zhang, Peixin and Sun, Jun and Wang, Xingmei},
  journal={arXiv preprint arXiv:2402.07518},
  year={2024}
}

@inproceedings{ghimire2024coding,
  title={Coding with ai: How are tools like chatgpt being used by students in foundational programming courses},
  author={Ghimire, Aashish and Edwards, John},
  booktitle={International Conference on Artificial Intelligence in Education},
  pages={259--267},
  year={2024},
  organization={Springer}
}

@article{lu2024entropy,
  title={An Entropy-based Text Watermarking Detection Method},
  author={Lu, Yijian and Liu, Aiwei and Yu, Dianzhi and Li, Jingjing and King, Irwin},
  journal={arXiv preprint arXiv:2403.13485},
  year={2024}
}

@inproceedings{sun2023codemark,
  title={Codemark: Imperceptible watermarking for code datasets against neural code completion models},
  author={Sun, Zhensu and Du, Xiaoning and Song, Fu and Li, Li},
  booktitle={Proceedings of the 31st ACM Joint European Software Engineering Conference and Symposium on the Foundations of Software Engineering},
  pages={1561--1572},
  year={2023}
}

@article{boiko2023autonomous,
  title={Autonomous chemical research with large language models},
  author={Boiko, Daniil A and MacKnight, Robert and Kline, Ben and Gomes, Gabe},
  journal={Nature},
  volume={624},
  number={7992},
  pages={570--578},
  year={2023},
  publisher={Nature Publishing Group UK London}
}

@InProceedings{kirchenbauer_watermark_2023,
  title = 	 {A Watermark for Large Language Models},
  author =       {Kirchenbauer, John and Geiping, Jonas and Wen, Yuxin and Katz, Jonathan and Miers, Ian and Goldstein, Tom},
  booktitle = 	 {Proceedings of the 40th International Conference on Machine Learning},
  pages = 	 {17061--17084},
  year = 	 {2023},
  series = 	 {Proceedings of Machine Learning Research},
  publisher =    {PMLR},
}

@inproceedings{christ_undetectable_2023,
  title={Undetectable watermarks for language models},
  author={Christ, Miranda and Gunn, Sam and Zamir, Or},
  booktitle={The Thirty Seventh Annual Conference on Learning Theory},
  pages={1125--1139},
  year={2024},
  organization={PMLR}
}

@inproceedings{atallah_natural_2001,
	location = {Berlin, Heidelberg},
	title = {Natural Language Watermarking: Design, Analysis, and a Proof-of-Concept Implementation},
	isbn = {978-3-540-45496-0},
	doi = {10.1007/3-540-45496-9_14},
	series = {Lecture Notes in Computer Science},
	shorttitle = {Natural Language Watermarking},
	pages = {185--200},
	booktitle = {Information Hiding},
	publisher = {Springer},
	author = {Atallah, Mikhail J. and Raskin, Victor and Crogan, Michael and Hempelmann, Christian and Kerschbaum, Florian and Mohamed, Dina and Naik, Sanket},
	editor = {Moskowitz, Ira S.},
	date = {2001},
        year = {2001},
	langid = {english},
}

@article{fu_gumbelsoft_2024,
  title={GumbelSoft: Diversified Language Model Watermarking via the GumbelMax-trick},
  author={Fu, Jiayi and Zhao, Xuandong and Yang, Ruihan and Zhang, Yuansen and et al.},
  journal={arXiv preprint arXiv:2402.12948},
  year={2024}
}

@inproceedings{koike_outfox_2024,
  title={Outfox: Llm-generated essay detection through in-context learning with adversarially generated examples},
  author={Koike, Ryuto and Kaneko, Masahiro and Okazaki, Naoaki},
  booktitle={Proceedings of the AAAI Conference on Artificial Intelligence},
  pages={21258--21266},
  year={2024}
}

@article{li2023starcoder,
  title={Starcoder: may the source be with you!},
  author={Li, Raymond and Allal, Loubna Ben and Zi, Yangtian and Muennighoff, Niklas and Kocetkov, Denis and Mou, Chenghao and Marone, Marc and Akiki, Christopher and Li, Jia and Chim, Jenny and others},
  journal={arXiv preprint arXiv:2305.06161},
  year={2023}
}

@article{guo2024deepseek,
  title={DeepSeek-Coder: When the Large Language Model Meets Programming--The Rise of Code Intelligence},
  author={Guo, Daya and Zhu, Qihao and Yang, Dejian and Xie, Zhenda and Dong, Kai and Zhang, Wentao and Chen, Guanting and Bi, Xiao and Wu, Yu and Li, YK and others},
  journal={arXiv preprint arXiv:2401.14196},
  year={2024}
}

@inproceedings{zhong_neural_2020,
	location = {Online},
	title = {Neural Deepfake Detection with Factual Structure of Text},
	url = {https://aclanthology.org/2020.emnlp-main.193},
	doi = {10.18653/v1/2020.emnlp-main.193},
	eventtitle = {{EMNLP} 2020},
	pages = {2461--2470},
	booktitle = {Proceedings of the 2020 Conference on Empirical Methods in Natural Language Processing ({EMNLP})},
	publisher = {Association for Computational Linguistics},
	author = {Zhong, Wanjun and Tang, Duyu and Xu, Zenan and Wang, Ruize and et al.},
	urldate = {2024-03-18},
	date = {2020-01-01},
        year = {2020},
}

@inproceedings{partialPython,
  title = {{{CoCoNuT}}: Combining Context-Aware Neural Translation Models Using Ensemble for Program Repair},
  shorttitle = {{{CoCoNuT}}},
  booktitle = {Proceedings of the 29th {{ACM SIGSOFT International Symposium}} on {{Software Testing}} and {{Analysis}}},
  author = {Lutellier, Thibaud and Pham, Hung Viet and Pang, Lawrence and Li, Yitong and et al.},
  year = {2020},
  month = jul,
  series = {{{ISSTA}} 2020},
  pages = {101--114},
  publisher = {{Association for Computing Machinery}},
  doi = {10.1145/3395363.3397369},
  url = {https://doi.org/10.1145/3395363.3397369},
  urldate = {2022-12-06},
}

@inproceedings{lee_who_2024,
    title = "Who Wrote this Code? Watermarking for Code Generation",
    author = "Lee, Taehyun et al.",
    booktitle = "Proceedings of the 62nd Annual Meeting of the Association for Computational Linguistics (Volume 1: Long Papers)",
    month = aug,
    year = "2024",
    publisher = "Association for Computational Linguistics",
    url = "https://aclanthology.org/2024.acl-long.268",
    pages = "4890--4911"
}

@article{yang2023watermarking,
  title={Watermarking text generated by black-box language models},
  author={Yang, Xi and Chen, Kejiang and Zhang, Weiming and Liu, Chang and Qi, Yuang and Zhang, Jie and Fang, Han and Yu, Nenghai},
  journal={arXiv preprint arXiv:2305.08883},
  year={2023}
}

@article{pan2024markllm,
  title={MarkLLM: An Open-Source Toolkit for LLM Watermarking},
  author={Pan, Leyi and Liu, Aiwei and He, Zhiwei and Gao, Zitian and et al.},
  journal={arXiv preprint arXiv:2405.10051},
  year={2024}
}

@article{chen2021codex,
  title={Evaluating large language models trained on code},
  author={Chen, Mark and Tworek, Jerry and Jun, Heewoo and et al.},
  journal={arXiv preprint arXiv:2107.03374},
  year={2021}
}

@article{guan2024codeip,
  title={Codeip: A grammar-guided multi-bit watermark for large language models of code},
  author={Guan, Batu and Wan, Yao and Bi, Zhangqian and Wang, Zheng and et al.},
  journal={arXiv preprint arXiv:2404.15639},
  year={2024}
}

@article{dishonesty,
title = {ChatGPT for good? On opportunities and challenges of large language models for education},
journal = {Learning and Individual Differences},
year = {2023},
issn = {1041-6080},
doi = {https://doi.org/10.1016/j.lindif.2023.102274},
url = {https://www.sciencedirect.com/science/article/pii/S1041608023000195},
author = {Enkelejda Kasneci and Kathrin Sessler and Stefan Küchemann and Maria Bannert et al.},
}

@article{hull,
author = {Ji, Ziwei and Lee, Nayeon and Frieske, Rita and Yu, Tiezheng and Su, Dan and Xu, Yan and Ishii, Etsuko and Bang, Ye Jin and Madotto, Andrea and Fung, Pascale},
title = {Survey of Hallucination in Natural Language Generation},
year = {2023},
issue_date = {December 2023},
publisher = {Association for Computing Machinery},
issn = {0360-0300},
url = {https://doi.org/10.1145/3571730},
doi = {10.1145/3571730},
journal = {ACM Comput. Surv.},
month = {mar},
articleno = {248},
}

@article{fairoze2023publicly,
  title={Publicly detectable watermarking for language models},
  author={Fairoze, Jaiden and Garg, Sanjam and Jha, Somesh and Mahloujifar, Saeed and Mahmoody, Mohammad and Wang, Mingyuan},
  journal={arXiv preprint arXiv:2310.18491},
  year={2023}
}

@inproceedings{
hu2024inevitable,
title={Inevitable Trade-off between Watermark Strength and Speculative Sampling Efficiency for Language Models},
author={Zhengmian Hu and Heng Huang},
booktitle={NeurIPS},
year={2024},
url={https://openreview.net/forum?id=6YKMBUiIsG}
}

\end{document}